\documentclass{article}

\usepackage[top=3cm, bottom=3cm, left=3cm,right=3cm]{geometry}
\usepackage{todonotes}
\usepackage{graphicx}
\usepackage{amssymb}
\usepackage{amsmath}
\usepackage{todonotes}
\usepackage{pdflscape}
\usepackage{caption}
\usepackage{subcaption}
\usepackage[sort,round]{natbib}
\usepackage[T1]{fontenc}
\usepackage[utf8]{inputenc}
\usepackage{authblk}
\usepackage{gensymb}

\def\ci{\perp\!\!\!\perp}

\title{Data Integration Model for Air Quality: A Hierarchical  Approach to  the Global  Estimation of Exposures to Ambient  Air Pollution}
\author[i]{Gavin Shaddick}
\author[i]{Matthew L. Thomas}
\author[i]{Amelia Jobling}
\author[ii]{Michael Brauer}
\author[iii]{Aaron van Donkelaar}
\author[iv]{Rick Burnett}
\author[v]{Howard H. Chang}
\author[vi]{Aaron Cohen}
\author[vii]{Rita Van Dingenen}
\author[viii]{Carlos Dora}
\author[viii]{Sophie Gumy}
\author[ix]{Yang Liu}
\author[iii]{Randall Martin}
\author[v]{Lance A. Waller}
\author[x]{Jason West}
\author[xi]{James V. Zidek}
\author[viii]{Annette Pr\"uss-Ust\"un}

\affil[i]{Department of Mathematical Sciences, University of Bath, U.K.}
\affil[ii]{School of Population and Public Health, The University of British Columbia, Canada}
\affil[iii]{Department of Physics and Atmospheric Science, Dalhousie University, Canada}
\affil[iv]{Health Canada, Ottawa, Canada}
\affil[v]{Department of Biostatistics and Bioinformatics, Rollins School of Public Health, Emory University, Atlanta, U.S.A.}
\affil[vi]{Health Effects Institute, Boston, U.S.A.}
\affil[vii]{Institute for Environment and Sustainability, Joint Research Centre, European Commission, Italy}
\affil[viii]{World Health Organization, Geneva, Switzerland}
\affil[ix]{Department of Environmental Health, Rollins School of Public Health, Emory University, Atlanta, U.S.A.}
\affil[x]{Department of Environmental Sciences and Engineering, University of North Carolina, Chapel Hill, U.S.A.}
\affil[xi]{Department of Statistics, University of British Columbia, Canada}

\begin{document}

\maketitle

\begin{abstract}
Air pollution is a major risk factor for global health, with both ambient and household air pollution contributing substantial components of the overall global disease burden. One of the key drivers of adverse health effects is fine particulate matter ambient pollution (PM$_{2.5}$) to which an estimated 3 million deaths can be attributed annually. The primary source of information for estimating exposures has been measurements from ground monitoring networks but, although coverage is increasing, there remain regions in which monitoring is limited. Ground monitoring data therefore needs to be supplemented with information from other sources, such as satellite retrievals of aerosol optical depth and chemical transport models. A hierarchical modelling approach for integrating data from multiple sources is proposed allowing spatially-varying relationships between ground measurements and other factors that estimate air quality. Set within a Bayesian framework, the resulting Data Integration Model for Air Quality (DIMAQ) is used to estimate exposures, together with associated measures of uncertainty, on a high resolution grid covering the entire world. Bayesian analysis on this scale can be computationally challenging and here approximate Bayesian inference is performed using Integrated Nested Laplace Approximations. Model selection and assessment is performed by cross-validation with the final model offering substantial increases in predictive accuracy, particularly in regions where there is sparse ground monitoring, when compared to previous approaches: root mean square error (RMSE) reduced from 17.1 to  10.7, and population weighted RMSE from 23.1 to 12.1 $\mu$gm$^{-3}$. Based on summaries of the posterior distributions for each grid cell, it is estimated that  92\% of the world's population reside in areas exceeding the World Health Organization's Air Quality Guidelines. 
\end{abstract}


\newpage

\section{Introduction}\label{sec:intro}

Ambient air pollution poses a significant threat to global health and has been associated with a range of adverse health effects, including cardiovascular and respiratory diseases in addition to some cancers \citep{brook2010particulate, sava2012respiratory, world2013, hoek2013long, newby2014expert, loomis14behalf}. Fine particulate matter (PM$_{2.5}$) in particular has been established as a key driver of global health with an estimated 3 million deaths in 2014 being attributable to PM$_{2.5}$ \citep{who2016}. It has been estimated that the majority of the world's population (87\%) reside in areas in which the World Health Organization (WHO) air quality guideline (annual mean of 10 $\mu$gm$^{-3}$) for PM$_{2.5}$ is exceeded \citep{brauer2015exposure}.  \\  

It is vital that the subsequent risks, trends and consequences of air pollution are monitored and modelled to develop effective  environmental and public health policy to lessen the burden of air pollution. Accurate measurements of exposure in any given area are required but this is a demanding task as, in practice, the processes involved are extremely complex and because of the scarcity of  ground monitoring data in some regions. The locations of ground monitoring sites within the WHO Air Pollution in Cities database \citep{world2016cities} are shown in Figure \ref{map:gm} where it can be seen that the density of monitoring sites varies considerably, with extensive measurements  available in North America, Europe, China and India but with little or no measurement data available for large areas of Africa, South America and the Middle East. \\

For this reason,  there is a need to use information from other sources in order to obtain estimates of exposures for all areas of the world. In 2013, the Global Burden of Disease study (\citet{forouzanfar2015global}, henceforth referred to as GBD2013) used a regression calibration approach to utilise information from satellite remote sensing and chemical transport models to  create a set of estimates of exposures on a high--resolution grid ($0.1^{\degree} \times 0.1^{\degree}$, approximately 11km $\times$ 11km at the equator) that were then matched to population estimates to estimate disease burden.  In GBD2013, a fused  estimate of PM$_{2.5}$, calculated as the average of estimates from satellites and chemical
transport models, was calibrated against ground measurements using linear regression.  For cells that contained a ground monitor,  measurements were regressed against this fused estimate in conjunction with information related to  local monitoring networks \citep{brauer2015exposure}. The resulting calibration function was applied to all grid cells, allowing a comprehensive set of global estimates of  PM$_{2.5}$ to be produced.\\

This allowed data from the three sources to be utilised, but the use of a single, global, calibration function resulted in underestimation in a number of  areas \citep{brauer2015exposure}. In reality,  the relationships between ground measurements and estimates from other sources will vary spatially due to regional differences in biases and errors that will be present in the different methods of estimation. Recently, \citet{van2016global} extended this approach using geographically weighted regression (GWR) to allow calibration (between the measurements and estimates) equations to vary spatially and to utilise additional information related to land use and the chemical composition of particulate matter.  However, both the original linear regression  and GWR approaches only provide an informal analysis of the uncertainty associated with the resulting estimates of exposure.  \\

In addition to regional differences in calibration functions, additional challenges arise when combining data  that are generated in fundamentally different ways.   Satellite pixels and chemical transport model cells are not the same with each potentially not capturing  different micro-scale features that may be reflected in the ground measurements and  all three sources of data will have different error structures that may not align. The difference in  resolution between ground monitors (point locations) and estimates from satellite and chemical transport models (grid cells) has led to the use of spatially varying coefficient models, often referred to as \emph{downscaling models} \citep{chang2016}. In the purely spatial model presented in \citet{berrocal2010spatio} for example, the intercepts and coefficients  are assumed to arise from continuous bivariate spatial process. Downscaling/upscaling models, set within a Bayesian hierarchical framework, have been used for both spatial and spatio-temporal modelling of air pollution with examples including  \citet{guillas2006statistical},  who used the UIUC 2-D chemical-transport model of the global atmosphere; \citet{van2006statistical}, who modelled  PM$_{10}$ concentrations over Western Europe using  information from both satellite observations and a chemical transport model; \citet{mcmillan2010combining}  who modelled PM$_{2.5}$ in the North Eastern U.S. using estimates from the Community Multi-scale Air Quality (CMAQ) numerical model; \citet{kloog2014new} who modelled PM$_{2.5}$ in the Northeastern U.S. using satellite-based aerosol optical depth (AOD) and  \citet{berrocal2010spatio} and \citet{zidek2012combining} who modelled ozone in the Eastern U.S. (Eastern and Central in the case of Zidek {\em et al.})  using estimates from CMAQ and a variant of the MAQSIP (Multiscale Air Quality Simulation Platform) model respectively. \\

In this setting, downscaling  is the calibration  of ground measurements with estimates obtained from other sources, e.g. satellites and chemical transport models, with random effects used to allow variation within grid-cells.  An alternative approach is that of \emph{Bayesian melding}  \citep{poole2000inference} in which both the measurements and the estimates are assumed to arise from an underlying latent process  that represents the true level of the pollutant. This latent process itself is unobservable but measurements can be taken, possibly with error,  at locations in space and time. For example,  the underlying latent process  represents the true level of PM$_{2.5}$ and this gives rise to the measurements from ground monitors and the estimates from satellite remote sensing and atmospheric models, all of which  will inform  the posterior distribution of the underlying latent process. Bayesian melding has been used to model SO$_2$ in the Eastern U.S. combining ground measurements with information from the Models-3 air quality model \citep{fuentes2005model}. Whilst Bayesian melding offers a flexible and coherent framework in which different models and processes can be combined, it is complex to implement and can be very computationally demanding, particularly using Markov chain Monte Carlo (MCMC), due to the requirement to perform a stochastic integral of the underlying continuous process to the resolution of the grid cells, for each grid cell. In contrast, one of the major advantages of downscaling is the computational saving that is made by only considering grid cells containing measurement locations within the estimation, after which   prediction at unknown locations is relatively straightforward \citep{chang2016}. However, unlike Bayesian melding, there is an implicit assumption the covariates are error free,   an assumption that  may be  untenable in practice.\\

In this paper, a model is presented for integrating data from multiple sources, that enables accurate estimation of global exposures to fine particulate matter. Set within a Bayesian hierarchical framework, this Data Integration Model for Air Quality (DIMAQ)  estimates exposures, together with associated measures of uncertainty, at high geographical resolution  by  utilising information from multiple sources and  addresses many of the issues encountered with previous approaches. The structure of the paper is as follows: after this introduction, Section 2 provides details of the data that are available, including measurements from ground monitoring and estimates from satellites and chemical transport models. Section 3 provides details of the statistical model (DIMAQ) that is used to integrate data from these different sources and  methods for  inference when performing Bayesian analysis with  large datasets. In Section 4 the results of applying  DIMAQ are presented, including examples of global and country specific estimates of exposure to PM$_{2.5}$ together with details of the methods used for model evaluation and comparison. Section 5 contains a  summary and a discussion of potential areas for  future research and Section 6 provides concluding remarks.

\section{Data} \label{sec:data}
 The sources of data used here can be allocated to one of three groups: (i) ground monitoring data; (ii)    estimates of PM$_{2.5}$ from remote sensing satellites and chemical transport models; (iii) other sources including  population,   land--use and topography.   Ground monitoring is available at a distinct number of locations, whereas the  latter  two groups provide near complete global coverage (and  have previously been shown to have strong associations  with global concentrations of PM$_{2.5}$, see below for details). Utilising such data  will allow
 estimates of exposures to be made for all areas, including those for which ground monitoring is sparse or non-existent.

\subsection{Ground measurements}

\begin{figure}[!b]
 \centering
 \includegraphics[scale=0.5]{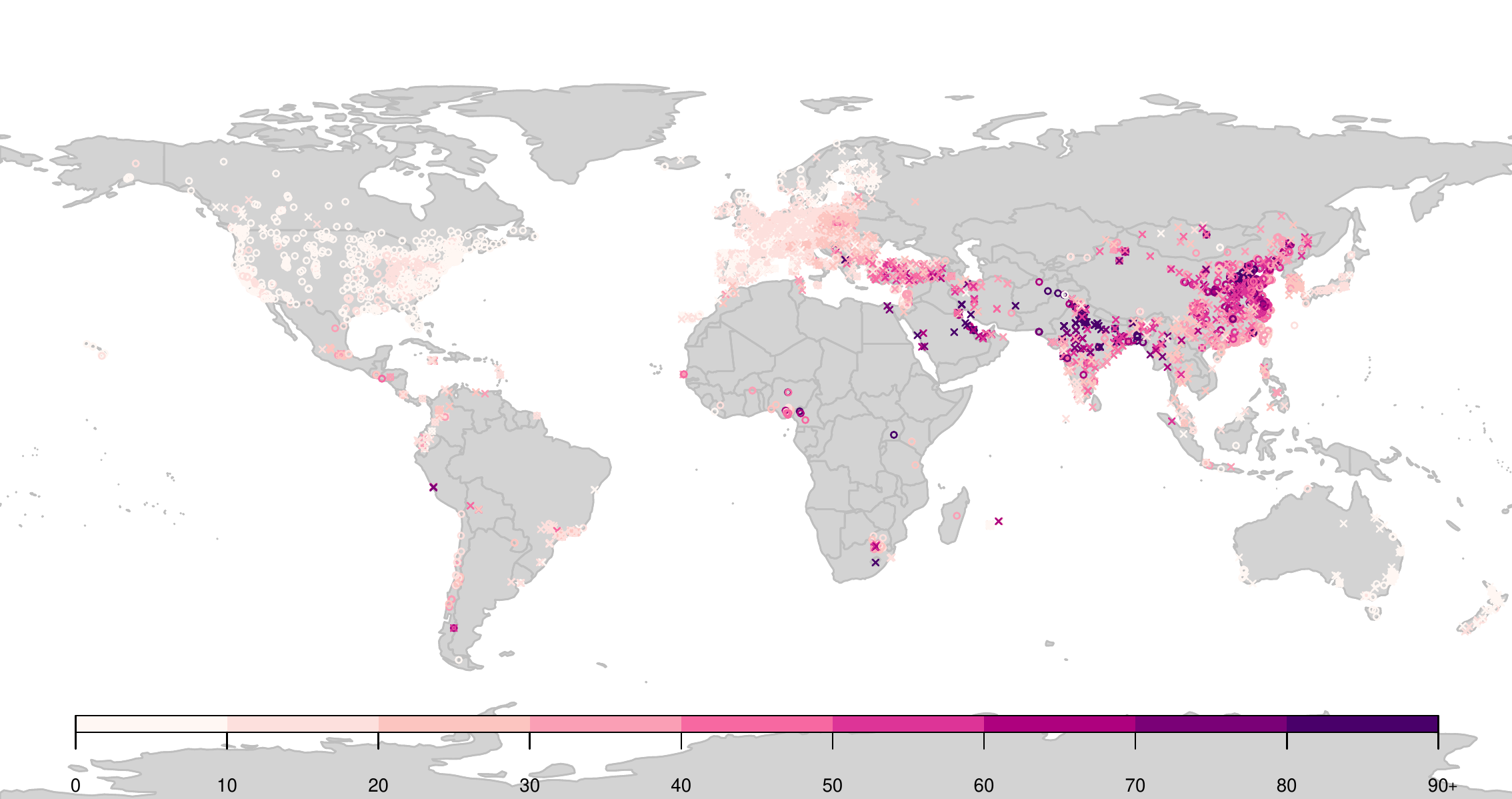}	
 \caption{Locations of ground monitors measuring PM$_{2.5}$ (circles) and PM$_{10}$ (crosses). Colours denote the annual average concentrations ($\mu$gm$^{-3}$) of PM$_{2.5}$ (or PM$_{2.5}$ converted from PM$_{10}$). Data are from 2014 (46\%), 2013 (36\%), 2012 (9\%) and 2006-2011, 2015 (9\%).} \label{map:gm}
 \end{figure}

 Ground measurements were available for locations reported within the WHO Air Pollution in Cities database \citep{world2016cities}, but rather than using the city averages reported in that database, monitor-specific measurements are used. The result was measurements of concentrations of PM$_{10}$ and PM$_{2.5}$ from 6,003 ground monitors. The locations and annual average concentrations for these monitors can be seen in Figure \ref{map:gm}. The database was compiled to represent measurements in 2014 with the majority of measurements coming from that year (2760 monitors). Where data were not available for 2014, data were used from 2015 (18 monitors), 2013 (2155), 2012 (564), 2011 (60), 2010 (375), 2009 (49), 2008 (21) and 2006 (1). In addition to annual average concentrations, additional information related to the ground measurements was also included where available, including monitor geo coordinates and monitor site type.\\
 
For locations measuring only PM$_{10}$, PM$_{2.5}$ measurements were estimated from PM$_{10}$. This was performed using a locally derived conversion factor (PM$_{2.5}$/PM$_{10}$ ratio, for stations where measurements are available for the same year) that was estimated using population-weighted averages of location-specific conversion factors for the country as detailed in \citet{brauer2012exposure}. If country-level conversion factors were not available, the average of country-level conversion factors within a region were used. 
 
 \subsection{Satellite-based estimates}
Satellite remote sensing is a method that estimates pollution from satellite retrievals  of aerosol optical depth (AOD), a measurement of light extinction by aerosols in the atmosphere. AOD indicates how aerosols modify the radiation leaving the top of the atmosphere after being scattered by the Earth's atmosphere and surface.  Estimates of PM$_{2.5}$ are obtained by correcting AOD using a spatially varying term, $\eta$,
\begin{equation*}
\mbox{PM}_{2.5} = \eta \times \mbox{AOD} \;.
\end{equation*}
Here $\eta$ is the coincident ratio of PM$_{2.5}$ to AOD and accounts for local variation in vertical structure, meteorology, and aerosol type. This ratio  is simulated from the GEOS-Chem  global chemical transport model \citep{bey2001global}. \\
 
\begin{figure}
 \centering
 \includegraphics[scale=0.3]{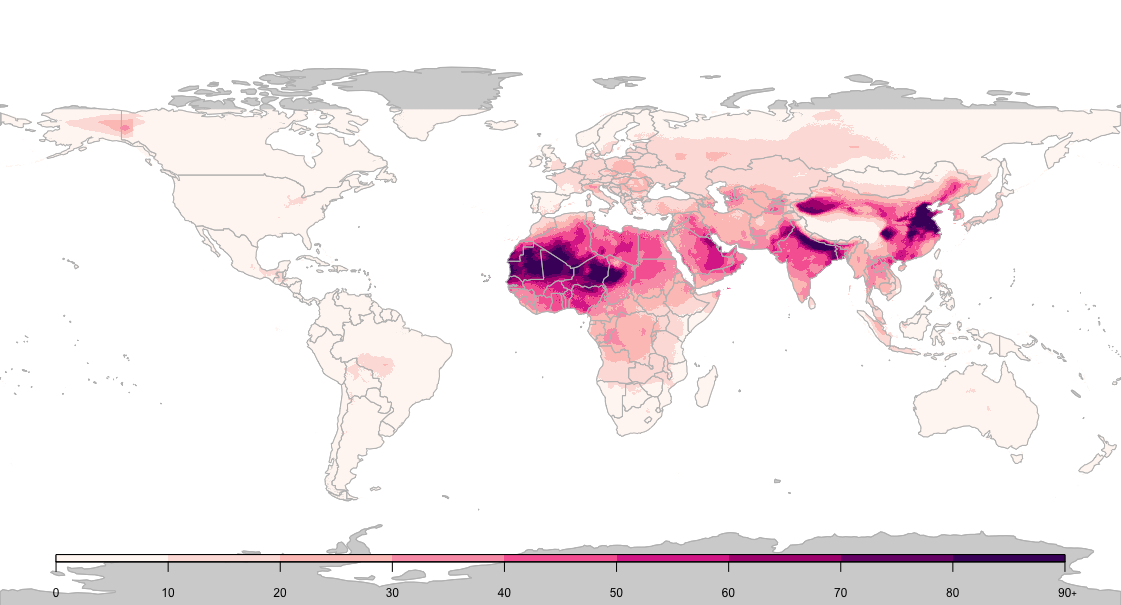}	
 \caption{Satellite based estimates of PM$_{2.5}$ ($\mu$gm$^{-3}$) for 2014, by grid cell ($0.1^{\degree}\times 0.1^{\degree}$ resolution).} \label{map:sat}
\end{figure}

The estimates used here combine AOD retrievals from multiple satellites with simulations from the GEOS-Chem chemical transport model and land use information, produced at a spatial resolution of  $0.1^{\degree} \times 0.1^{\degree}$, which is approximately 11km $\times$ 11km at the equator. This is described in detail in \citet{van2016global}. A map of the estimates of PM$_{2.5}$ from this model can be seen in Figure \ref{map:sat}.
 
\subsection{Chemical transport model simulations}
  
\begin{figure} 
 \centering
 \begin{subfigure}{\textwidth}  \centering
 \includegraphics[width=0.7\linewidth]{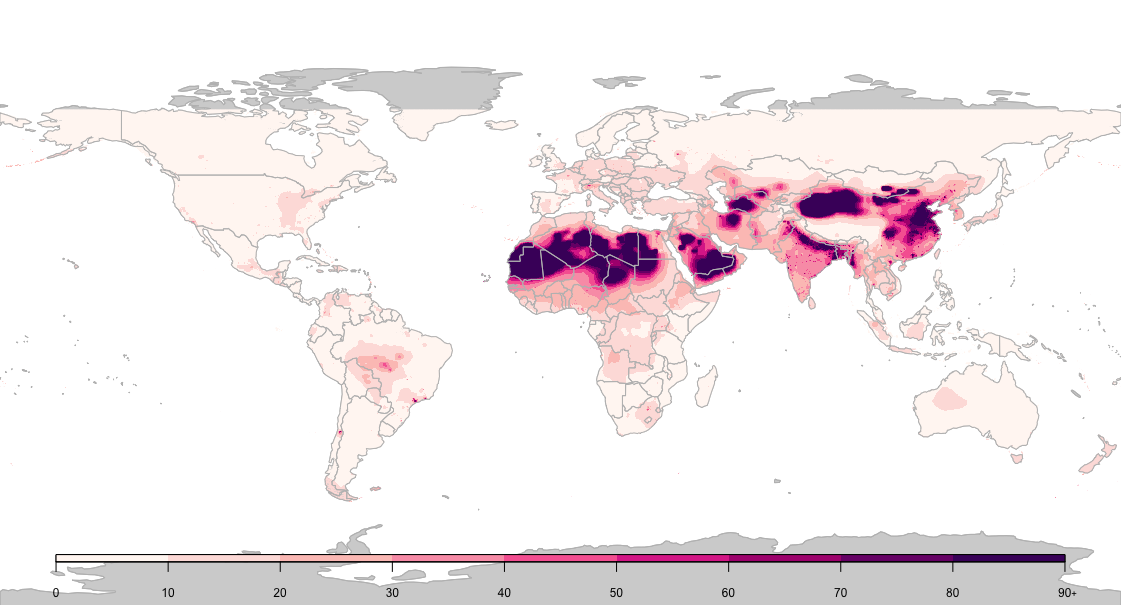}	
 \caption{Estimates of PM$_{2.5}$ ($\mu$gm$^{-3}$)  for 2010 from the TM5 chemical transport model used in GBD2013.} \label{map:ctm1}
 \end{subfigure}
  \begin{subfigure}{\textwidth}  \centering
 \includegraphics[width=0.7\linewidth]{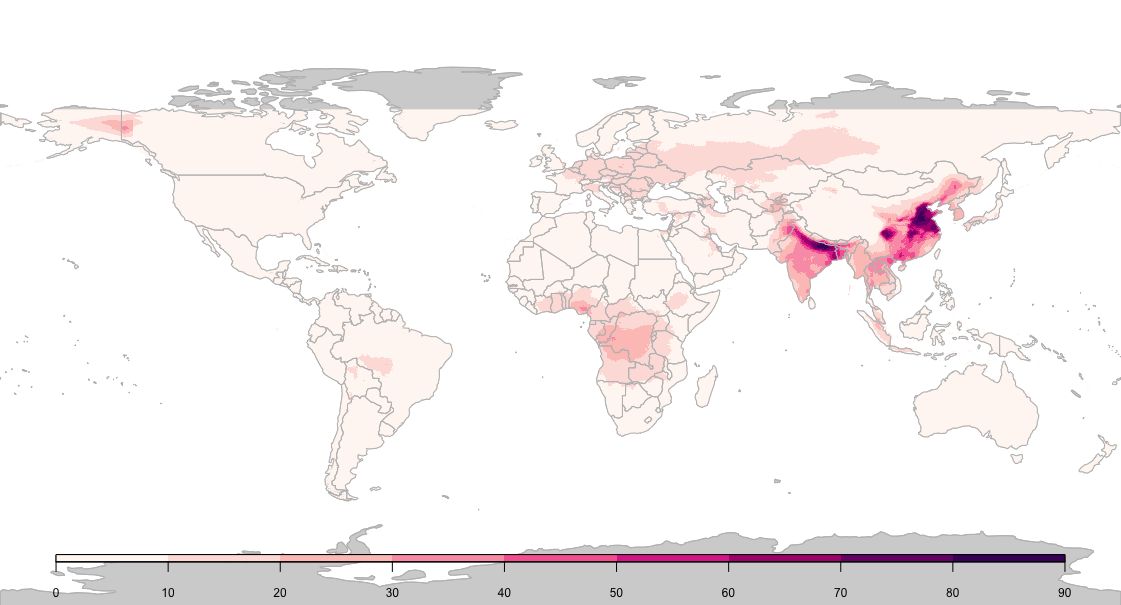}	
 \caption{Estimates of the sum of sulphate, nitrate, ammonium and organic carbon  for 2014 from the GEOS-Chem chemical transport model.} \label{map:ctm2}
 \end{subfigure}
   \begin{subfigure}{\textwidth} 
 \centering
 \includegraphics[width=0.7\linewidth]{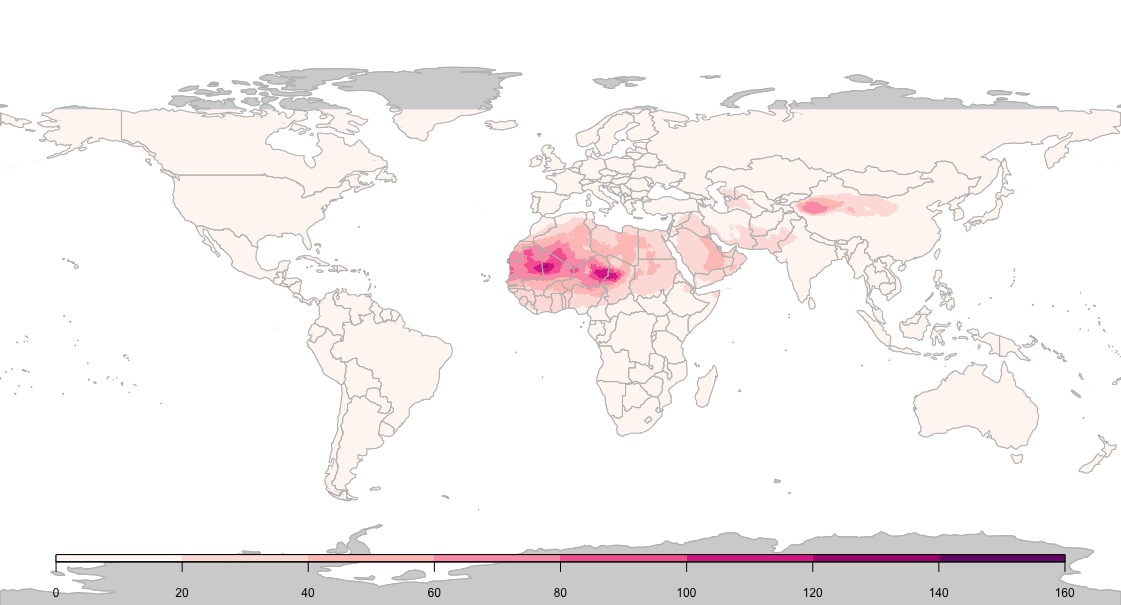}	
 \caption{Estimates of the compositional concentrations of mineral dust  for 2014 from the GEOS-Chem chemical transport model.} \label{map:ctm3}
 \end{subfigure}  
 \caption{Estimates from chemical transport models, by grid cell ($0.1^{\degree}\times 0.1^{\degree}$ resolution).}
 \end{figure}

Numerically simulated estimates of PM$_{2.5}$  were obtained  from atmospheric chemical transport models. There are a variety of such models available including GEOS-Chem \citep{bey2001global}, TM5 \citep{huijnen2010global} and  TM5-FASST \citep{van2014multi}. The first two of these are nested 3-dimensional global atmospheric transport models which can be used to simulate levels of PM$_{2.5}$ with TM5-FASST being a reduced form of the full TM5 model, developed to allow faster computation for impact assessment \citep{van2014multi}. Estimates at a spatial resolution of  $1^{\degree} \times 1^{\degree}$ were allocated to a higher resolution grid, of   $0.1^{\degree} \times 0.1^{\degree}$, based on population density \citep{brauer2012exposure, brauer2015exposure}. Estimates for PM$_{2.5}$  from the TM5-FASST model  were available  for 2010, as described  in \citet{brauer2015exposure}. A map of these estimates can be seen in Figure \ref{map:ctm1}.\\

In addition to the estimates of PM$_{2.5}$,  estimates of the sum of sulphate, nitrate, ammonium and organic carbon (SNAOC) and the compositional concentrations of mineral dust  (DUST) based on simulations from the GEOS-Chem chemical transport model \citep{van2016global} were available  for 2014. Maps of the estimates of SNAOC and DUST can be seen in Figures \ref{map:ctm2} and \ref{map:ctm3} respectively.

\subsection{Population data}
\begin{figure}
 \centering
 \includegraphics[scale=0.3]{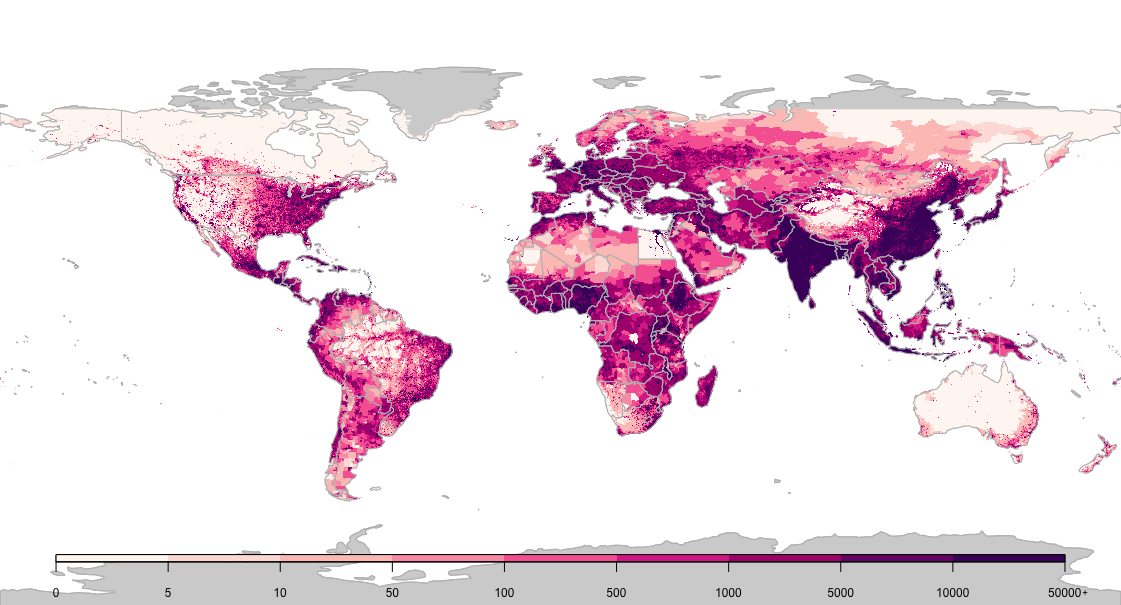}	
 \caption{Population estimates for 2014 from the Gridded Population of the World version 4 (GPW v4)  database, by grid cell ($0.1^{\degree}\times 0.1^{\degree}$ resolution).} \label{map:pop}
\end{figure}

A comprehensive set of population data on a high-resolution grid was obtained from the Gridded Population of the World (GPW) database. These data are provided on a $0.0417^{\degree}\times 0.0417^{\degree}$ resolution. Aggregation to each $0.1^{\degree}\times 0.1^{\degree}$ grid cell was performed as detailed in \citet{brauer2015exposure}. GPW version 4 provides population estimates for 2000, 2005, 2010, 2015 and 2020. Following the methodology used in \citet{brauer2015exposure}, populations for 2014 were obtained by interpolation using cubic splines (performed for each grid cell) with knots placed at 2000, 2005, 2010, 2015 and 2020. A map of the resulting estimates of populations for 2014 can be seen in  Figure \ref{map:pop}.
  
\subsection{Land use}

\citet{van2016global} developed a measure combining  information on elevation and land use that was shown to be a significant predictor of PM$_{2.5}$. 
For each ground monitor, the following are calculated: (i) the difference between the elevation (of the ground monitor) and that of the surrounding  grid cell, as defined by the GEOS-Chem chemical transport model (ED); (ii) the distance to the nearest urban land surface (DU), based upon  MODIS land cover descriptions \citep{friedl2010modis}. The resulting measure (ED $\times$ DU) was available  for 2014 for each $0.1^{\degree} \times 0.1^{\degree}$ grid cell.

\section{Statistical Modelling} \label{sec:statsmod}
The aim  is to obtain estimates of concentrations of PM$_{2.5}$ for each of  1.4 million grid cells, together with associated measures of uncertainty. This will be achieved by finding the posterior distributions for each cell, from which summary measures will be calculated. \\

The overall approach is statistical calibration as described in  \citet{chang2016}:  a regression model is used to express ground measurements, $Y_s$,   available at a discrete set of $N_S$ locations $S \in \cal{S}$ with  labels $S= \{s_0,s_1,\dots,s_{N_S}\}$, that are  a function of covariates, $X_{sr}$: $r=1,\dots, R$, that reflect  information from other sources,  as described in Section \ref{sec:data}. Covariate information may be available for point locations (as with the ground measurements)  or on a grid of $N_L$  cells, $l \in L$  where $L= l_1, ..., l_{N_L}$. \\

Considering a single covariate, $X_{lr}$, for ease of explanation,

\begin{equation} \label{eqn:3}
Y_s = \tilde{\beta}_{0s} + \tilde{\beta}_{1s} X_{lr} + \epsilon_s 
\end{equation}
where $X_{lr}$ is measured on a grid. Here,  $\epsilon_s \sim N(0,\sigma_\epsilon^2)$ is a random error term. The terms $\tilde{\beta}_{0s}$ and $\tilde{\beta}_{1s}$ denote random effects that allow the intercept and coefficient to vary over space
\begin{eqnarray*}
\tilde{\beta}_{0s} = \beta_0+ \beta_{0s} \\
\tilde{\beta}_{1s} = \beta_1+ \beta_{1s} \; .
\end{eqnarray*}
Here, $\beta_0$ and $\beta_1$ are fixed effects: representing the mean value of the intercept and coefficients respectively, with $\beta_{0s}$ and $\beta_{1s}$ zero mean spatial random effects: providing (spatially driven) adjustments to these means, allowing the calibration functions to vary over space. In downscaling models, it is assumed that the parameters $\beta_{0s}$ and $\beta_{1s}$ arise from a  continuous  spatial process which allows within grid-cell variation (see \citet{berrocal2010spatio} for an example using a continuous bivariate spatial process). \\

Despite  monitoring data being increasingly available, there are  issues that may mean using a spatial continuous process may be problematic in this setting. Monitoring protocols,  measurement techniques, quality control procedures and mechanisms for obtaining annual averages may vary from country-to-country \citep{brauer2012exposure} leading to natural discontinuities in ground measurements, and their precision, between countries. In addition, the geographic distribution of measurements, as seen in Figure \ref{map:gm}, is heavily biased toward North America, Europe, China and India with some areas of the world, e.g. Africa, having  very little monitoring information to inform such a model.  \\

\begin{figure}
 \centering
 \begin{subfigure}{\textwidth}  \centering
 \includegraphics[width=0.8\linewidth]{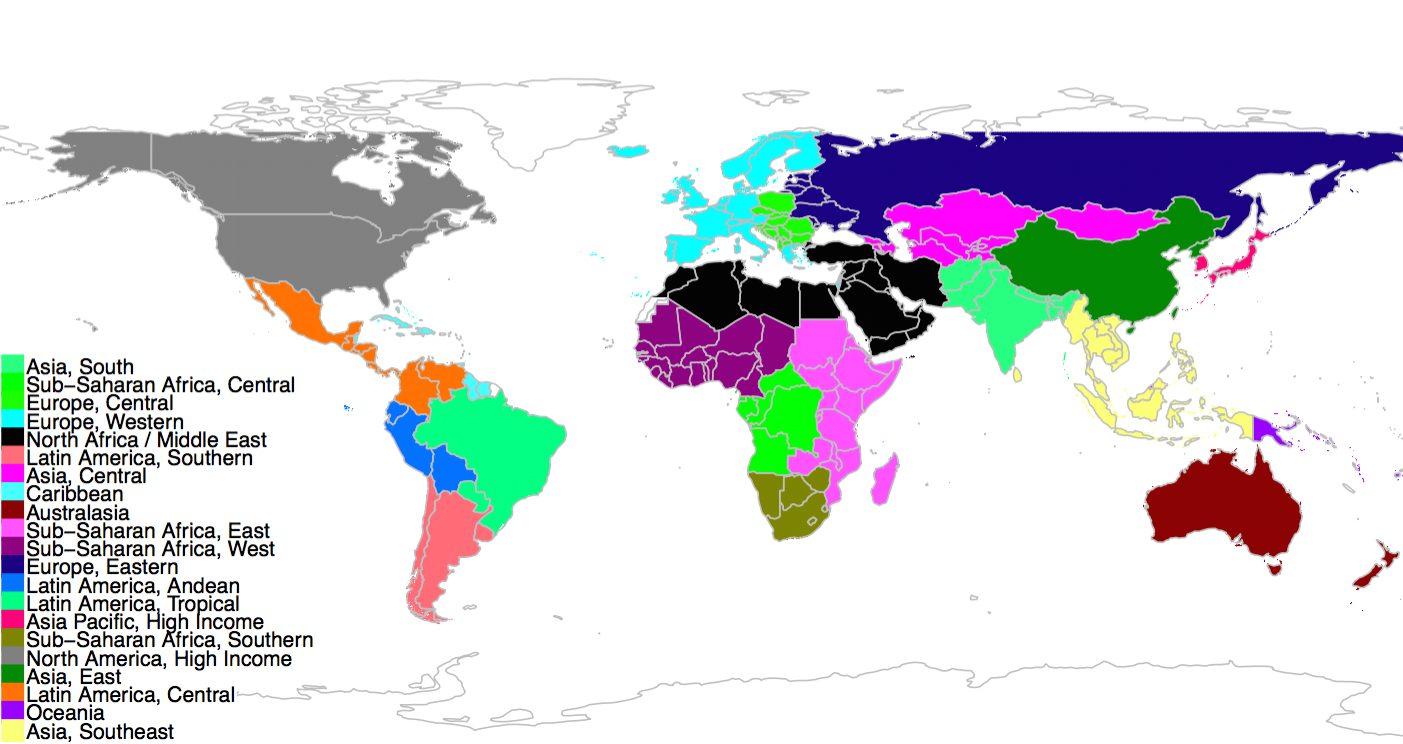}	
 \caption{Map of regions.}
 \end{subfigure}
  \begin{subfigure}{\textwidth}
 \centering
 \includegraphics[width=0.8\linewidth]{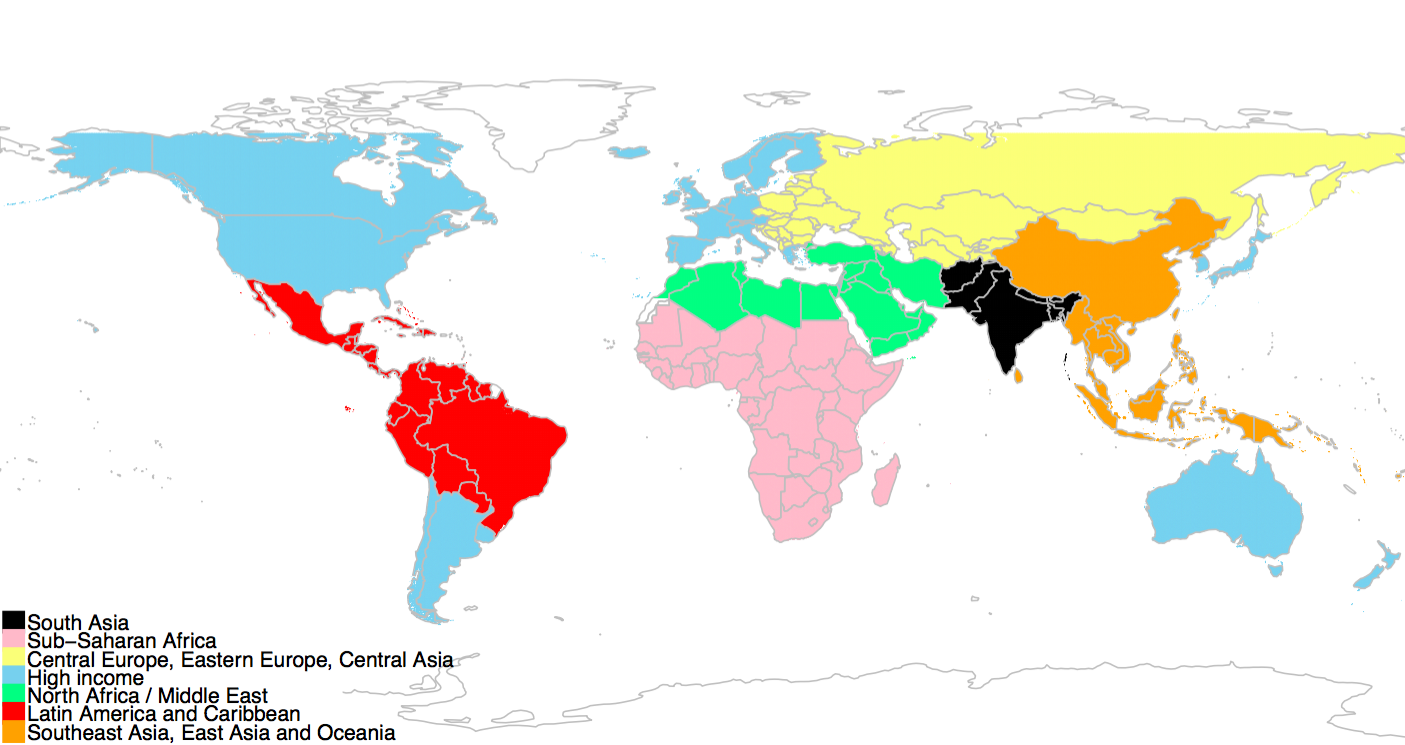}	
 \caption{Map of super-regions.}
 \end{subfigure}
 \caption{Schematic showing the nested geographical structure of countries within regions within super-regions. } \label{map:regions}
 \end{figure}

The structure of the random effects used here exploits a geographical nested hierarchy: each of the 187 countries considered are  allocated to one of 21 regions and, further, to one of 7 super-regions. Each region must contain  at least two countries and  is broadly based on geographic regions/sub-continents and groupings based on country level development status and causes of death \citep{brauer2012exposure}. The geographical structure of regions within super-regions can be seen in Figure \ref{map:regions}. Where there are limited  monitoring data within a country,  information can be borrowed from  higher up  the hierarchy, i.e. from other countries within the region and further, from  the wider super-region. It is noted that the `high income' super-region is non-contiguous and for North Africa/Middle East  the region is the  same as the super-region and therefore will be a single set of random effects, i.e. no distinction between  region and super-region, for this area. \\

\subsection{A Data Integration Model for Air Quality}\label{sec:DIMAQ}

Annual averages of ground measurements (of PM$_{2.5}$) at point locations, $s$, within grid cell, $l$, country, $i$, region, $j$,  and  super--region, $k$  are denoted by $Y_{slijk}$. As described in Section \ref{sec:statsmod}, there is a nested hierarchical structure with $s=1,\dots, N_{lijk}$ sites within grid cell, $l$; $l=1,\dots, N_{ijk}$, grid cells within country $i$;  $i=1,\dots, N_{jk}$, countries within region  $j$; $j=1,\dots, N_k$,  regions  within super--region $k$: $k=1,\dots, N$.  In order to allow for the  skew in the measurements  and the constraint of non--negativity, the (natural) logarithm of the measurements are used.  \\

The model  consists of  sets of fixed and random effects, for both intercepts and covariates, and is given as follows,

\begin{eqnarray} \label{eqn:arr1} \nonumber
\log(Y_{slijk}) = \tilde{\beta}_{0,lijk} &+& \sum_{q \in Q} \tilde{\beta}_{q, ijk} X_{q,lijk} \\ \nonumber 
&+& \sum_{p_1 \in P_1} \beta_{p_1} X_{p_1,lijk} + \sum_{p_2 \in P_2} \beta_{p_2} X_{p_2,slijk} \\
&+& \epsilon_{slijk} \;, 
\end{eqnarray}
where $\epsilon_{slijk} \sim N(0,\sigma_{\epsilon}^2)$ is a random error term.   A set of $R$ covariates contains two groups, $R=(P, Q)$, where $P$ are those which have fixed effects (across space) and $Q$ those  assigned random effects. The main estimates of air quality, e.g. those from satellites and chemical transport models, will be assigned random effects and are in $Q$, with other variables being assigned fixed effects. Within the group, $P$,  of covariates that have fixed effects;  $P_1$ are  available at the grid cell level, $l$, with others, $P_2$, being available for the point locations, $s$, of the monitors, $P=(P_1, P_2)$. 

\subsubsection{Structure of the random effects}

Here,  the random effect terms, $\tilde{\beta}_{0,lijk}$ and $\tilde{\beta}_{q,ijk}$, have contributions from  the country, the region and the super--region, with the intercept also having a random effect for the cell representing within-cell variation in ground measurements,  
\begin{eqnarray} \nonumber
\tilde{\beta}_{0, lijk} &=& \beta_0+ \beta_{0,lijk}^G + \beta_{0,ijk}^C + \beta_{0,jk}^R +\beta_{0,k}^{SR} \\ \nonumber
\tilde{\beta}_{q, ijk} &=& 	 \beta_q+  \beta_{q,ijk}^C + \beta_{q,jk}^R +\beta_{q,k}^{SR} \; .
\end{eqnarray}
For clarity of exposition, the following description is restricted to a generic parameter, $\beta$. Let $\beta^{SR}_k$ denote the coefficient for  super-region $k$. The coefficients for super-regions  are distributed with mean equal to the overall mean ($\beta_0$, the fixed effect) and variance, $\sigma_{SR}^2$, representing between super-region variability,
\begin{equation*}
\beta^{SR}_{k} \sim N(\beta_0,\sigma_{SR}^2) \,
\end{equation*}
where $k=1,\ldots,N = 7$. Similarly, each super-region contains a number of regions. Let $\beta^{R}_{jk}$ denote the coefficient for region $j$ (in super--region $k$) that will be distributed with mean equal to to the coefficient for the  super-region and variance representing the between region (within super--region) variability. 
\begin{equation*}
\beta^{R}_{jk} \sim N(\beta^{SR}_k,\sigma^2_{R,k}) 
\end{equation*}
where $j=1,\ldots,N_k$,  the number of regions in super-region $j$. Each region will  contain a number of countries. Let $\beta^{C}_{ijk}$ denote the coefficient for  country $i$ in region $j$ and super-region $k$. The country level effect will be distributed with mean equal  to the coefficient for region $j$ within super-region $k$ with variance representing  the between country (within region) variability.
\begin{equation} \label{eqn:countr1}
\beta^{C}_{ijk} \sim N(\beta^{R}_{jk},\sigma^2_{C, jk})
\end{equation}
where $i=1,\ldots,N_{jk}$ is   the number of countries in region $j$ (in super--region $k$). Note that in the case of the intercepts, there is an additional term, $\beta^G_{lijk}$,  representing  within grid cell (between monitoring locations) variability. \\

Country effects within regions and regional effects within super--regions  are assumed to be independent  within their respective geographies. However, the geographical hierarchy is broadly based on geographic regions, sub-continents, mortality and economic factors \citep{brauer2012exposure} and, as such, there are countries for which the allocation may not be optimal  when considering environmental factors, such as air pollution. For example, Mongolia is included within the Asia Central region and Central Eastern Europe and Central Asia super--region (see Figure \ref{map:regions}) but its pollution profile might be expected to be more similar to those of its  direct neighbours, including China  (which is in a different region (Asia East) and super--region (South East Asia, East Asia and Oceania),  than the profiles of more western countries. For this reason, it might be advantageous to allow the borrowing of information in Equation (\ref{eqn:countr1}) to include countries that are immediate neighbours rather than all of the countries in the surrounding administrative region. This could be achieved using an intrinsic conditionally autoregressive (ICAR) model \citep{besag1974spatial} in place of Equation (\ref{eqn:countr1}), 
\begin{equation} \label{eqn:countr2} \nonumber
\beta^{C}_{i} | \beta^{C}_{i'}, \; i' \in \partial_i \sim N\left(
  \overline{\beta}^{C}_{i},\frac{\psi^2}{N_{\partial i}}\right),
  \end{equation}
where $\partial i$ is the set of neighbours of country $i$, $N_{\partial i}$ is the number of neighbours, and $\overline{\beta}^{C}_i$ is the mean of the spatial random effects of these neighbours.

\subsection{Inference}

The model presented in Section \ref{sec:DIMAQ} is a Latent Gaussian Model (LGM) which means that advantage can be taken of methods offering  efficient computation when performing Bayesian inference. LGMs can be implemented using approximate Bayesian inference using integrated nested Laplace Approximations (INLA) as proposed in \citet{rue2009approximate} using the R-INLA software \citep{rue2012r}.  The following sections provide a brief  summary of LGMs (Section \ref{sec:LGM}) and INLA  (Section \ref{sec:INLA}) with additional details linking to the  model described in Section \ref{sec:DIMAQ}.

\subsubsection{Latent Gaussian models} \label{sec:LGM}
  The  model presented in Equation \ref{eqn:arr1} can be expressed in general form as follows: Given $\eta_s = g(E(Y_s))$, where  $g()$ is a link function, 
  \begin{eqnarray} \nonumber
\eta_s &=& \beta_0 + \sum_{p=1}^{P} \beta_p X_{qs} + \sum^Q_{q=1} f_q (Z_{qs}) 	
\end{eqnarray}

\noindent where $\beta_0$ is an overall intercept term, the set of $\beta_p$ $(p=1,\dots, P)$ are the coefficients associated with covariates, $X$; the fixed effects. The set of functions,  $f_1(\cdot), \dots, f_Q(\cdot)$ represent the random effects with the form of the function being determined by the model. For example a hierarchical model may have, $f_1(\cdot) \sim N(0, \sigma^2_f)$, with a distribution defined for $\sigma^2_f$, whereas for standard regression, $f(\cdot) \equiv 0$, leaving just fixed effects. \\

The set of unknown parameters, $\theta$, will include both the coefficients of the model shown above and the parameters required for the functions, i.e. $\theta = (\beta_p, f_q)$. Here $\theta$ will contain the parameters of the model as described in Section \ref{sec:DIMAQ} and will include  $\beta_0, \beta_{0,lijk}^G, \beta_{0,ijk}^C, \beta_{0,jk}^R, \beta_{0,k}^{SR}, \beta_q,   \beta_{q,ijk}^C,  \beta_{q,jk}^R$ and $\beta_{q,k}^{SR}$, with the set of  hyperparameters associated with $\theta$ being $\psi_2=(\sigma^2_{G,i},\sigma^2_{C,j}, \sigma^2_{R,j}, \sigma^2_{SR})$. The overall set of parameters, $\psi=(\psi_1, \psi_2)$, also contains  $\psi_1=(\sigma^2_{\epsilon})$, which relates to the variance of the measurement error in the data. \\

Assigning a Normal distribution to the parameters in $\theta$,  $\theta | \psi \sim  MVN(\theta | \Sigma (\psi_2)$) will result in a LGM.  The computation required to perform inference will be largely determined by the characteristics of the covariance matrix, $\Sigma (\psi_2)$, which  will often be dense, i.e. it will have many entries that are non-zero, leading  to a high computational burden when performing  the matrix inversions that will be required to perform inference. If  $\theta | \psi_2$ can be expressed in terms of a Gaussian Markov Random Field (GMRF),  then it may be possible to take advantage of methods that reduce computation when performing Bayesian analysis on models of this type \citep{rue2005gaussian}. Using a GMRF means that   typically  the inverse of the covariance matrix, $Q = \Sigma^{-1}$ will be sparse, i.e. more zero entries,  due to the conditional independence between sets of parameters in which $\theta_l \ci  \theta_m | \theta_{-lm} \iff Q_{lm}=0$ (where $-lm$ denotes the vector of $\theta$ with the $l$ and $m$ elements removed) \citep{rue2005gaussian}. Expressing  $\theta | \psi_2$ in terms of the precision, rather than the covariance, gives $\theta | \psi \sim  MVN(\theta | Q(\psi_2)^{-1}$), where  $\psi_2$  denotes the parameters associated with $Q$ rather than $\Sigma$. \\

\subsubsection{Integrated Laplace Approximations}\label{sec:INLA}

Estimation of the (marginal) distributions of the model parameters and hyperparameters of a LGM  will require evaluation of the following integrals:

\noindent \begin{eqnarray} \nonumber \label{eqn:lgm2} 
	p(\theta_j | Y) &=& \int p(\theta_j | Y, \psi)p(\psi|Y) d \psi \\ 
		p(\psi_k | Y) &=& \int p(\psi | Y) d \psi_{-k}
\end{eqnarray}

In all but the most stylised cases, these will not be analytically tractable. Samples from these distributions could be obtained using Markov-chain Monte Carlo methods but  there may be issues when fitting  LGMs using MCMC, as described in \citet{rue2009approximate} and  the computational burden may be excessive, especially large numbers of predictions are required.   Here,   approximate Bayesian inference is performed  using INLA. It is noted that the dimension of $\theta$ is much larger than the dimension of $\psi$ and this will help in the implementation of the model as the computational burden increases linearly with the dimension of $\theta$ but exponentially with the dimension of $\psi$.\\

The aim is to find approximations for the distributions shown in Equation (\ref{eqn:lgm2}).  For the hyperparameters, the posterior of $\psi$ given $Y$ can be written as

\begin{eqnarray} \nonumber
p(\psi|Y) &=& \frac{p(\theta,\psi| Y)}{p(\theta| \psi, Y)} \\ \nonumber
&\propto & \frac{p(Y|\theta)p(\theta|\psi)p(\psi)}{p(\theta|\psi, Y)} \\  \nonumber
&\approx & \left.\frac{p(Y|\theta)(\theta|\psi)p(\psi)}{\tilde{p}(\theta|\psi, Y)} \right| _{\theta = \hat{\theta}(\psi)} \\  \nonumber
&=& \tilde{p}(\psi | Y)\; .
\end{eqnarray}
Here a Laplace approximation (LA) is used in the denominator for $\tilde{p}(\theta|\psi, Y)$. For univariate $\theta$ with an integral of the form $\int e^{g(\theta)}$, the LA takes the form $g(\theta) \sim N(\hat{\theta}(\psi), \hat{\sigma}^2)$, where $\hat{\theta}(\psi)$ is the modal value of $\theta$ for specific values of the hyperparameters, $\psi$ and $\hat{\sigma}^2 = \left\{ \frac{d^2 \log g(\theta)}{d \theta^2} \right \}^{-1}$. \\

The mode of $\tilde{p}(\psi | Y)$ can be found numerically by Newton type algorithms. Around the mode, the  distribution, $\log \tilde{p} (\psi | Y)$, is evaluated over a grid of $H$ points,  $\{\psi^*_h\}$, with  associated integration weights $\Delta_h$.  For each point on the grid, the marginal posterior, $\tilde{p}(\psi_h^* | Y)$ is obtained from which approximations to the marginal distributions,  $\tilde{p}(\psi |Y)$, can be found using numerical integration. \\

For the  individual model parameters, $\theta_j$,

\begin{eqnarray} \nonumber
p(\theta_j|Y) &=& \frac{p((\theta_j, \theta_{-j}),\psi| Y)}{p(\theta_{-j}| \theta_{j}, \psi, Y)} \\ \nonumber
&\propto & \frac{p(Y|\theta)p(\theta|\psi)p(\psi)}{p(\theta_{-j}| \theta_{j}, \psi, Y)}\\ \nonumber
&\approx & \left.\frac{p(Y|\theta)(\theta|\psi)p(\psi)}{\tilde{p}(\theta_{-j}| \theta_{j}, \psi, Y)} \right| _{\theta_{-j} = \hat{\theta_{-j}}(\theta_{j}, \psi)} \\ \nonumber
&=& \tilde{p}(\theta_j | \psi, Y)\; .
\end{eqnarray}

For an LGM, $(\theta_{-j}| \theta_{j}, \psi, Y)$ will be approximately Gaussian. There are a number of ways of constructing the approximation in the denominator including a simple Gaussian approximation which will be   computationally attractive but may be inaccurate. Alternatively, a LA would be  highly accurate but computationally expensive. R-INLA uses a computationally efficient method,  a simplified LA, that consists of performing a Taylor expansion around the LA of $\tilde{p}(\theta_j|\psi, Y)$, aiming to
`correct' the Gaussian approximation for location and
skewness \citep{rue2009approximate}.\\

The marginal posteriors, $\tilde{p}(\psi^*_h | Y)$, that have been evaluated at points $\{\psi^*_h\}$, are used to obtain the conditional posteriors,  $\tilde{p}(\theta_j | \psi^*_h, Y)$, on a grid of values for $\theta_j$. The marginal posteriors, $\tilde{p}(\theta_j |  Y)$, are then found by  numerical integration: $\tilde{p}(\theta_j |  Y) = \sum^H_h  \tilde{p}(\theta_j | \psi^*_h, Y)\tilde{p}(\psi^*_h|Y) \Delta_h$, with the integration weights, $\Delta_h$ being  equal when the grid takes the form of a regular lattice. \\

The model presented in Section \ref{sec:DIMAQ}  was implemented using  R-INLA  \citep{rue2012r} installed on the Balena high performance computing system (HPC) at the the University of Bath \newline \verb (www.bath.ac.uk/bucs/services/hpc/facilities/).  Fitting the model described in Section \ref{sec:DIMAQ} to data from the 6003 monitors (and associated covariates) does not itself require the use of an HPC but the prediction on the entire grid (of 1.4 million cells) did present some computational challenges. When fitting the model, the prediction locations are treated as missing data and their posterior distributions are approximated simultaneously with  model fitting.  INLA requires a copy of the model to be stored on a single  node which,  even with the high-memory compute nodes (32GBs per core) available with the HPC, resulted in memory issues when attempting to perform estimation and prediction on the entire grid in a single step. Therefore, prediction was performed using subsets of the prediction grid,  each containing groups of regions.  Each subset, including  the satellite estimates and other variables included in the model,   was appended to the modelling dataset with both estimation and prediction performed for each combination.  The resulting sets of predictions  were combined to give  a complete set of global predictions.

\section{Results}

A series of models based on the structure described in Section \ref{sec:DIMAQ} were applied with the aim of assessing the predictive ability of potential explanatory factors. The choice of which variables were included in the final model was made based on their contribution to within-sample model fit and out-of-sample predictive ability.  \\

In the comparisons that follow, model (i) is the model used in GBD2013 \citep{brauer2015exposure} and is a  linear regression model with response equal to the average concentration from monitors within a grid cell and covariates $X_1, X_2$ and $X_3$ together with the  average of the satellite based estimates and those from the TM5-FASST chemical transport model for each cell, $(X_4+X_5)/2$. Models (ii) to (v) are variants  of the model presented in Section \ref{sec:DIMAQ}.\\
 
Details of the variables included in five candidate models can be seen in Table \ref{tab:models}. They include information on local network characteristics; indicator variables for whether the type of monitor was unspecified, $X_1$; whether the exact location is known, $X_2$, and whether PM$_{2.5}$ was estimated from PM$_{10}$ ($X_3$); satellite based estimates of PM$_{2.5}$ concentrations ($X_4$),  estimates of PM$_{2.5}$ ($X_5$) from the TM5-FASST chemical transport model,  dust (DUST; $X_6$) and the sum of sulphate, nitrate, ammonium and organic carbon (SNAOC; $X_7$) from atmospheric models;  estimates of  population ($X_8$) and a function of land--use and elevation (ED $\times$ DU; $X_9$).  Except for the measurements themselves, all of these variables  are spatially aligned to the resolution of the  grid. Further details  can be found in Section \ref{sec:data}. \\

For evaluation, cross validation was performed using 25 combinations of training (80\%) and validation (20\%) datasets.  Validation sets were obtained by taking a stratified random sample, using sampling probabilities based on the cross-tabulation of PM$_{2.5}$ categories (0-24.9, 25-49.9, 50-74.9, 75-99.9, 100+ $\mu$gm$^{-3}$) and super-regions, resulting in concentrations in each validation sets having the same distribution of PM$_{2.5}$ concentrations and super-regions as the overall set of sites.  The following metrics were calculated for each training/evaluation set combination: for model fit: R$^{2}$ and deviance information criteria (DIC, a measure of model fit for Bayesian models) and for predictive accuracy:  root mean squared error (RMSE) and population weighted root mean squared error (PwRMSE).\\ 

\begin{table}[hbt]
\centering
  \begin{tabular}{ccccccccccc}
{\bf Model} &\multicolumn{2}{c}{\bf  (i)}  &\multicolumn{2}{c}{\bf  (ii)} 
 & \multicolumn{2}{c}{\bf  (iii)}  &\multicolumn{2}{c}{\bf  (iv)}
  &\multicolumn{2}{c}{\bf  (v)}\\
 Variable & F & R & F & R& F & R& F & R& F & R \\
    \hline
   Intercept &   \checkmark & &\checkmark&\checkmark&\checkmark&\checkmark&\checkmark&\checkmark&\checkmark&\checkmark \\
    $X_1^{\dagger}$ &   \checkmark & &\checkmark&&\checkmark&&\checkmark&&\checkmark\\
    $X_2^{\dagger}$ &   \checkmark & &\checkmark&&\checkmark&&\checkmark&&\checkmark\\
    $X_3^{\dagger}$ &   \checkmark & &\checkmark&&\checkmark&&\checkmark&&\checkmark\\
$X_4$ &   \checkmark & &\checkmark&\checkmark&\checkmark&\checkmark&\checkmark&\checkmark&\checkmark&\checkmark \\
$X_5$ &   \checkmark & &&&\checkmark&\checkmark&&&\checkmark&\checkmark \\
$X_6$ &   & &&&&&\checkmark&&\checkmark& \\
$X_7$ &   & &&&&&\checkmark&&\checkmark& \\
$X_8^{\dagger\dagger}$ &    & &\checkmark&\checkmark&\checkmark&\checkmark&\checkmark&\checkmark&\checkmark&\checkmark\\
$X_9$ &   & &&&&&\checkmark&&\checkmark& \\
    \hline\\
 \multicolumn{11}{l} {    $\dagger$ Together with interaction with $X_4$, $X_5$ where they are}\\  \multicolumn{11}{l} {  included within the model. }\\
\multicolumn{11}{l} { $\dagger\dagger$ Country level random  
effects are assigned a}\\
 \multicolumn{11}{l} {conditional autoregressive prior.}
  \end{tabular}
  \caption{Variables includes in each of five candidate models: $X_1$, whether the type of monitor was unspecified; $X_2$, whether the exact location is known; $X_3$, whether PM$_{2.5}$ was estimated from PM$_{10}$;  $X_4$, satellite based and $X_5$ chemical transport model estimates of PM$_{2.5}$; $X_6$ and $X_7$, estimates of compositional concentrations of mineral dust and the sum of sulphate, nitrate, ammonium and organic carbon from atmospheric models; $X_9$, a function of elevation difference and land--use.  Here, F and R denote the inclusion of fixed  and random effects respectively for each   variable.  } \label{tab:models}
\end{table}

\begin{figure}
 \centering
 \includegraphics[scale=0.5]{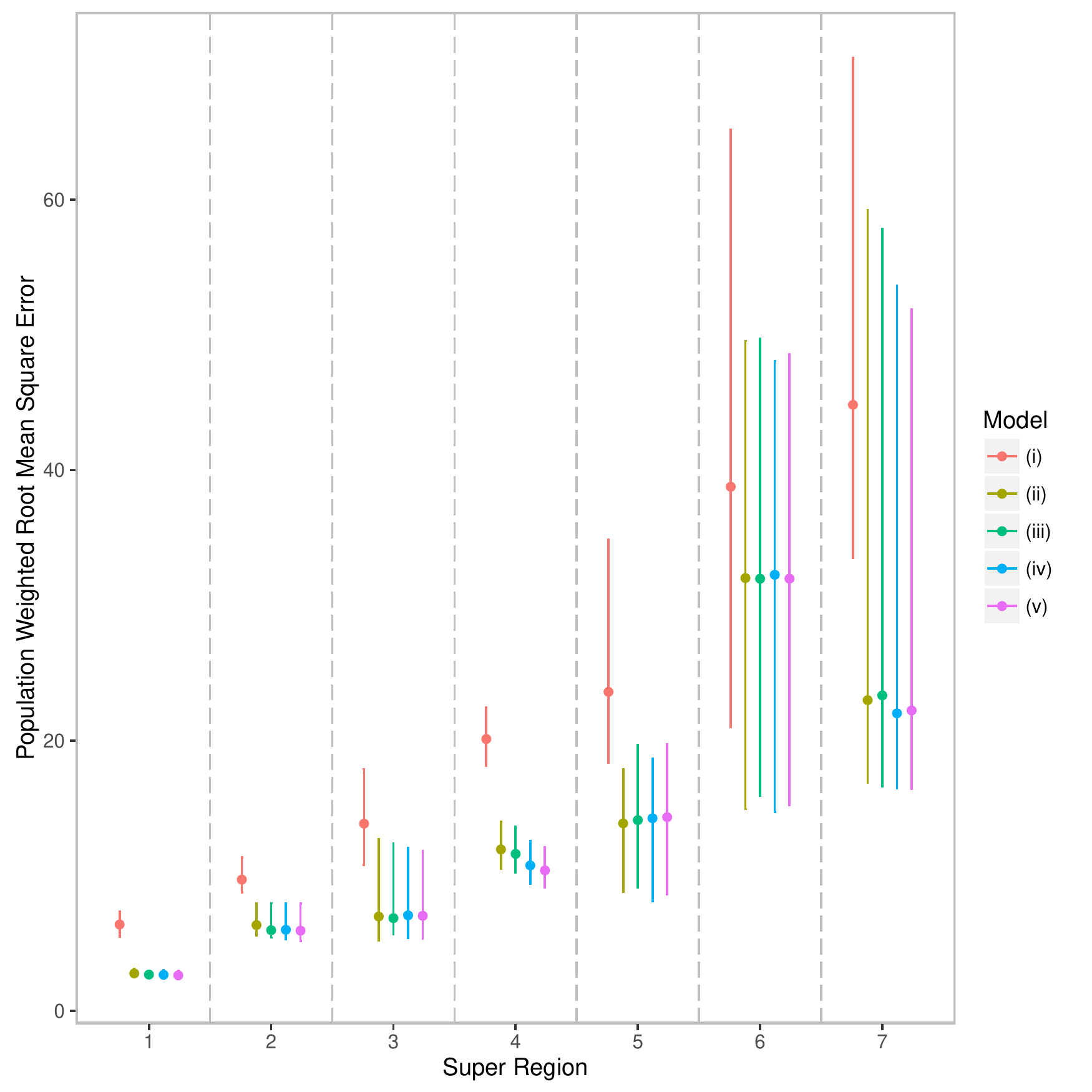}	
 \caption{Summaries of predictive ability of the GBD2013 model (i) and four candidate models (ii-iv), for each of seven super--regions: 1, high income; 2, Central Europe, Eastern Europe, Central Asia; 3, Latin America and Caribbean; 4, Southeast Asia, East Asia and Oceania; 5, North Africa / Middle East; 6, Sub-Saharan Africa; 7, South Asia. For each model,  population weighted root mean squared errors ($\mu$gm$^{-3}$) are given with dots denoting the median of the distribution from 25 training/evaluation sets and the vertical lines the range of values.} \label{fig:res1}
 \end{figure}

 The results of fitting the five candidate models can be seen in Table \ref{tab:tab1}, which shows     R$^2$ and DIC for within sample model fit and  RMSE and PwRMSE for out-of-sample predictive ability, and in Figure \ref{fig:res1} which shows the PwRMSE for each model by super-region. It can be seen that using any of the hierarchical models based on the structure described in Section \ref{sec:DIMAQ} provides an immediate improvement in all metrics when compared to the linear model, with a single global calibration function, used in GBD2013. For example, using model (ii) which contains satellite based estimates, population and local network characteristics  results in the overall R$^2$ improving from 0.54 to 0.90, DIC from 7828 to 1105 and reductions of  5.9 and 10.1 $\mu$gm$^{-3}$ for RMSE and population weighted RMSE respectively. This improvement can be seen in each of the super-regions (Figure \ref{fig:res1}), with the most marked improvements in areas where there is limited ground monitoring. \\  
 
 Adding either estimates of PM$_{2.5}$ from the TM5-FASST chemical transport model; model (iii), or estimates of specific chemical components (SNAOC) and dust (DUST) from the GEOS-Chem chemical transport model together with  information on differences in elevation between a ground monitor and its surrounding grid cell (ED $\times$ DU); model (iv), to this resulted in further improvements with model (iv) showing the most improvement. Although it resulted  in a reduction in the DIC, adding the estimates of PM$_{2.5}$ from the TM5-FASST chemical transport model to model (iv) did not result in any substantial improvement in  predictive ability. This may be in part due to the fact that the variables used in model (iv) are produced for 2014 whereas the estimates from the TM5-FASST model are from 2010. Considering the lack in improvement of predictive ability and the increased complexity and computational burden involved when incorporating an additional set of random effects, these estimates are not  not included in the final model (model (iv)).  \\

\begin{table}[h]
\centering
\begin{normalsize}
\begin{tabular}{clllll}
\multicolumn{1}{c} {\bf Model} & \multicolumn{1}{c} {\bf R$^2$} & \multicolumn{1}{c} {\bf DIC} & \multicolumn{1}{c} {\bf RMSE$\dagger$} & \multicolumn{1}{c} {\bf PwRMSE$\dagger$} \\
\hline 
(i) & 0.54 (0.53, 0.54)& 7828 (7685, 8657)& 17.1 (16.5, 18.1)&  23.1 (20.5, 29.3)\\
(ii) & 0.90 (0.90, 0.91)& 1105 (849, 1239)& 11.2 (10.1, 12.9)& 13.0 (11.5, 23.5)\\
 (iii) & 0.90 (0.90, 0.91)& 986 (704, 1115)& 11.1 (10.0, 13.3)& 12.8\ (11.2, 23.0)\\
 (iv) & 0.91 (0.90, 0.91)& 877 (640, 1015)& 10.7 (9.5, 12.3)& 12.1 (10.7, 21.4)\\
 (v) & 0.91 (0.90, 0.92)& 777 (508, 919) & 10.7 (9.5, 12.5)& 12.0 (10.7, 20.7)\\
\hline 
$\dagger \; \mu$gm$^{-3}$
\end{tabular}
\end{normalsize}
\caption{Summary of results from fitting five candidate models. Model  (i) is the model used in the GBD2013 study. Model (ii) is a   hierarchical model containing satellite based estimates of PM$_{2.5}$, population and  local network characteristics. Models (iii-v) contain  additional variables:  model (iii),     estimates  of PM$_{2.5}$ from   the TM5-FASST chemical transport  model;  model (iv),   estimates of specific chemical components and dust from the GEOS-Chem chemical transport model and information on differences in elevation between a ground monitor and its surrounding grid cell (as defined by the GEOS-Chem chemical transport model); (v)  both the estimates from the TM5-FASST and GEOS-Chem  models. Results are for both in-sample model fit and out-of-sample predictive ability and  are the median (minimum, maximum) values from 25 training-validation set combinations. For  within sample model fit,   R$^2$ and Deviance Information Criteria (DIC).  For out-of-sample predictive ability, root mean squared error (RMSE) and population weighted root mean squared error (PwRMSE). } \label{tab:tab1}
\end{table}

\begin{figure} 
 \centering
 \begin{subfigure}{\textwidth}  \centering
 \includegraphics[width=0.8\linewidth]{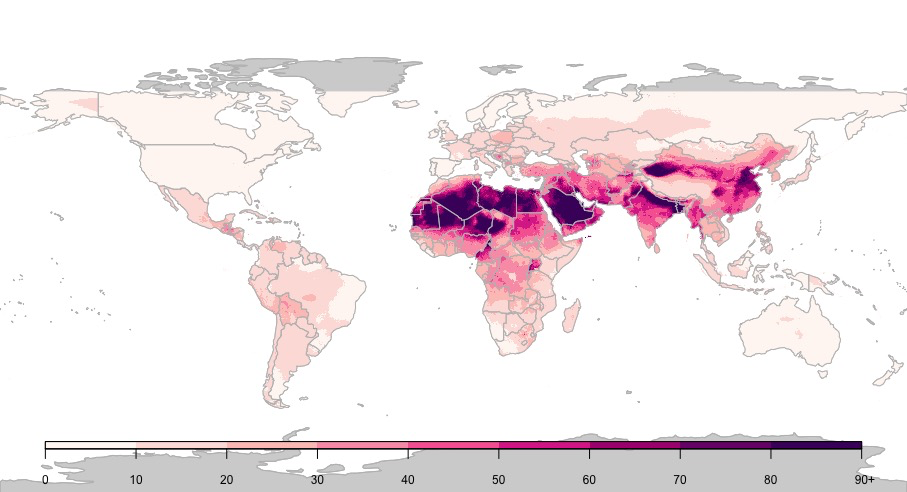}
 \caption{Medians of  posterior distributions.} 
 \label{map:pred}
 \end{subfigure}
  \begin{subfigure}{\textwidth}  \centering
 \includegraphics[width=0.8\linewidth]{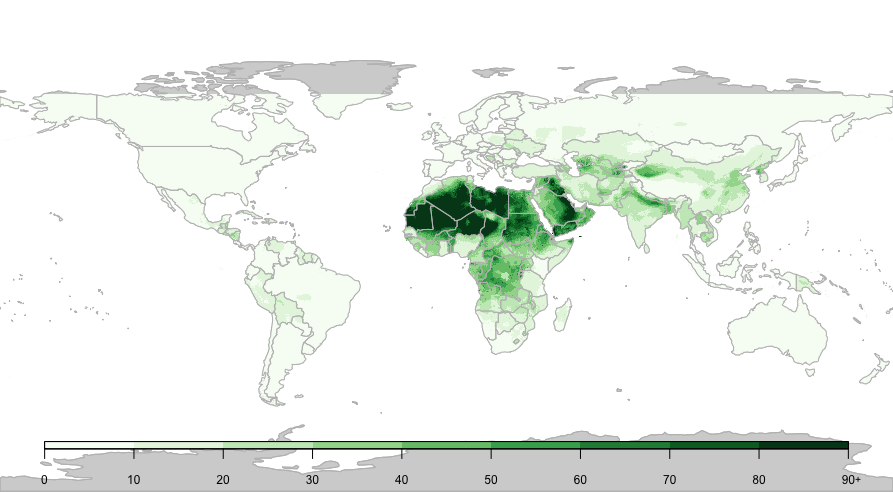}
 \caption{Half the width of  95\% posterior credible intervals.} \label{map:ci}
 \end{subfigure} 
 \caption{Estimates of annual averages of PM$_{2.5}$ ($\mu$gm$^{-3}$) for 2014 together with  associated uncertainty for each grid cell ($0.1^{\degree} \times 0.1^{\degree}$ resolution) using a Bayesian hierarchical model (see text for details).}
 \label{map:world}
 \end{figure}

Predictions from the final model (model (iv)) can be seen in Figures \ref{map:pred} and \ref{map:ci}. The point estimates shown in Figure \ref{map:pred} give a summary of air quality for each  grid cell and show clearly the spatial variation in global PM$_{2.5}$. For each grid cell, there is an underlying (posterior) probability distribution which incorporates information about the uncertainty of these estimates. There are a number of ways of presenting this uncertainty and Figure \ref{map:ci} shows one of these; half of the length of the 95\% credible intervals \citep{denby2007uncertainty}. Higher uncertainty will be combination of higher concentrations and sparsity of monitoring data and the effects of these can be clearly seen in areas of North Africa and the Middle East. \\
 
\begin{figure} 
 \centering
 \begin{subfigure}{\textwidth}  \centering
 \includegraphics[width=0.6\linewidth, clip=TRUE, trim = {0 10mm 0 18mm}]{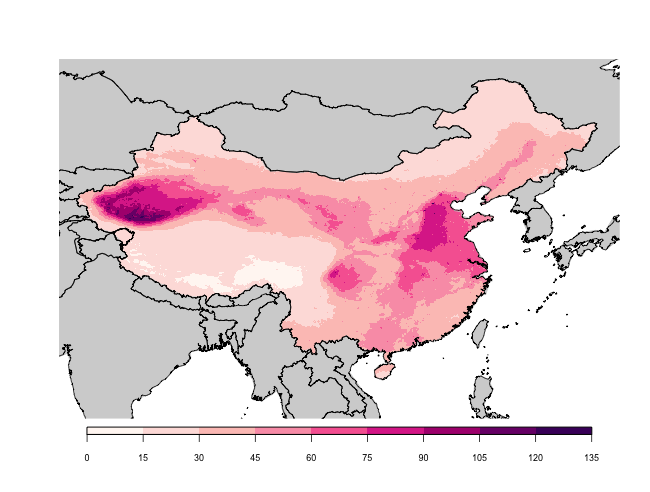}
 \caption{Medians of  posterior distributions.} 
\label{map:china1}
 \end{subfigure}
  \begin{subfigure}{\textwidth}  \centering
 \includegraphics[width=0.6\linewidth, clip=TRUE, trim = {0 10mm 0 18mm}]{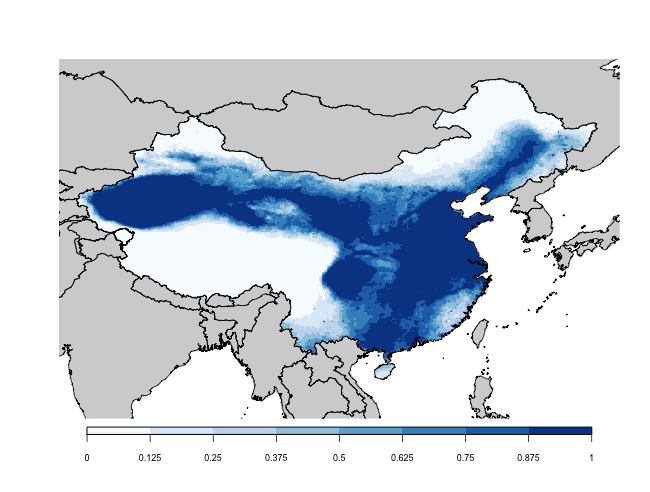}
 \caption{Probability of  exceeding 35 $\mu$gm$^{-3}$.} 
 \label{map:china2}
 \end{subfigure} 
  \begin{subfigure}{\textwidth}  \centering
 \includegraphics[width=0.6\linewidth, clip=TRUE, trim = {0 10mm 0 18mm}]{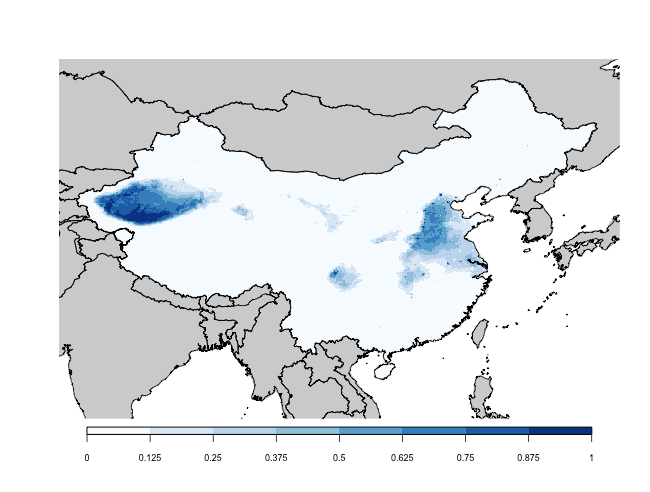}
 \caption{Probability of  exceeding 75 $\mu$gm$^{-3}$.}
 \label{map:china3}
 \end{subfigure} 
 \caption{Estimates of annual mean PM$_{2.5}$ concentrations ($\mu$gm$^{-3}$) for 2014 together with  exceedance probabilities using a Bayesian hierarchical model (see text for details) for each grid cell ($0.1^{\degree} \times 0.1^{\degree}$ resolution) in China.}
 \label{map:china}
 \end{figure}

\begin{figure} 
 \centering
 \begin{subfigure}{0.47\textwidth}  \centering
 \includegraphics[width=\linewidth]{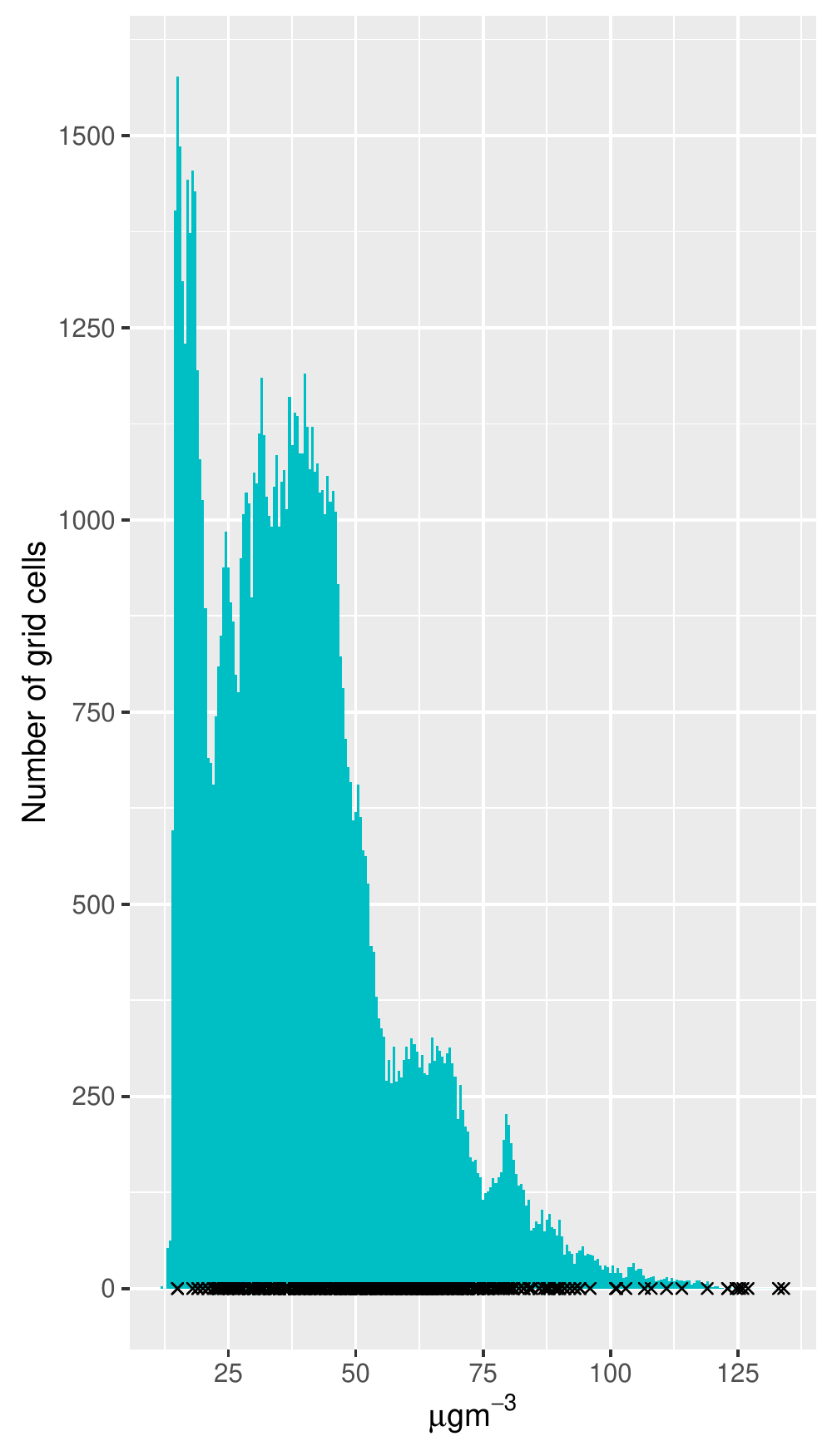}
 \caption{Estimated annual average concentrations of PM$_{2.5}$ by grid cell ($0.1^o \times 0.1^o$ resolution). Black crosses denote the annual averages recorded at ground monitors.}
\label{map:milliegram1}
 \end{subfigure}\hspace{5pt}
  \begin{subfigure}{0.47\textwidth}  \centering
 \includegraphics[width=\linewidth]{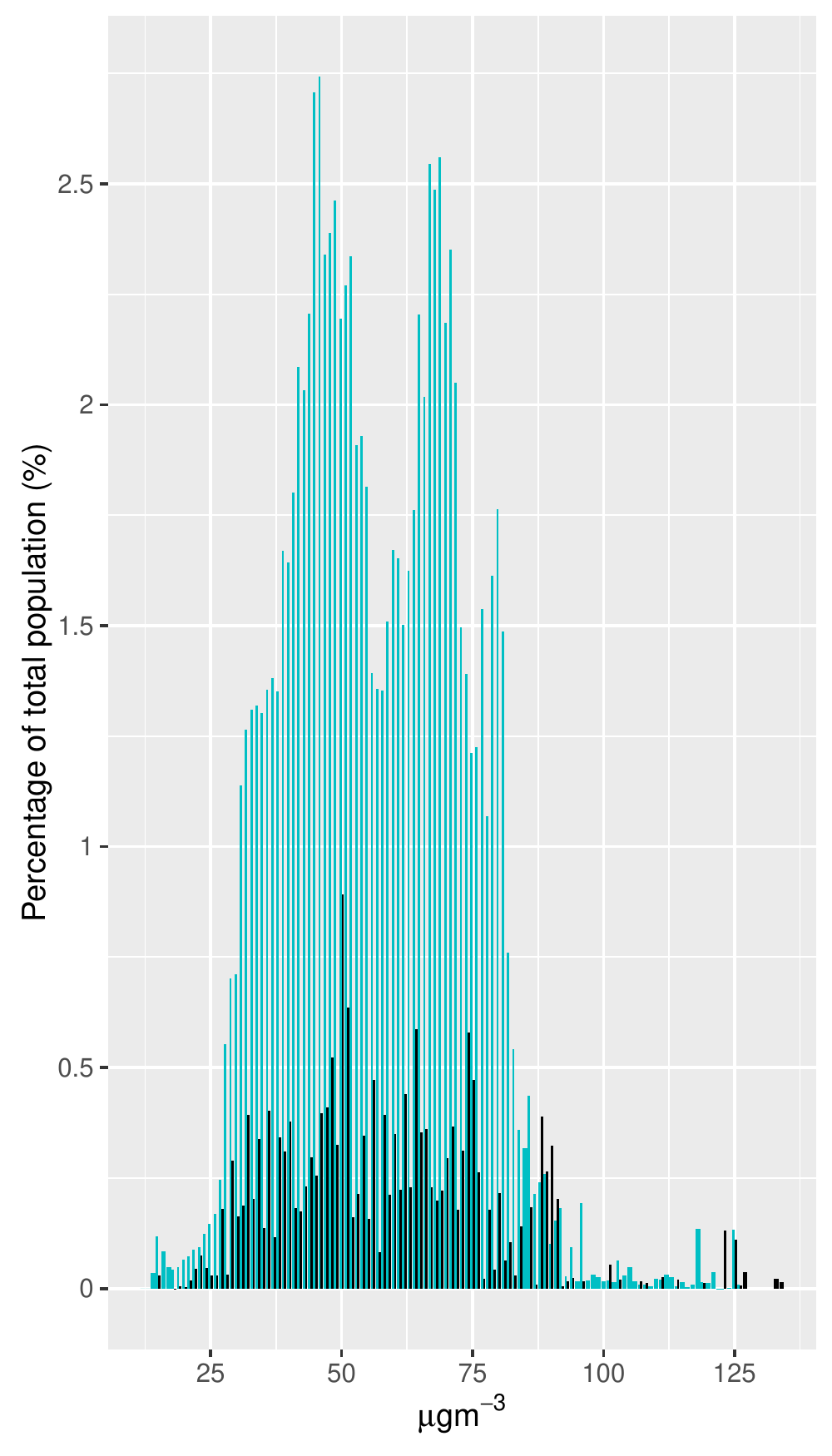}
 \caption{Estimated population level exposures (blue bars) and, for cells containing at least one monitoring station,   population weighted measurements from ground monitors (black bars). } 
 \label{map:milliegram2}
 \end{subfigure} 
 \caption{Distributions of annual mean concentrations  and population level exposures for PM$_{2.5}$ ($\mu$gm$^{-3}$) in China.}
 \label{map:milliegram}
 \end{figure}

 The distributions for each cell can also be used to examine the probabilities of exceeding particular thresholds. Figure \ref{map:china} shows an example of this and contains predicted concentrations for China (Figure \ref{map:china1}) together with the probability for each cell that the value exceeds 35 $\mu$gm$^{-3}$ (Figure \ref{map:china2}) and 75 $\mu$gm$^{-3}$ (Figure \ref{map:china3}). High probabilities of exceeding the greater of the two thresholds are observed in the area around Beijing and in the Xinjiang province in the far west of the country. For the latter,  a substantial component of the high (estimated) concentrations  will  be  due to mineral dust from the  large deserts in the region, as can be seen in Figure \ref{map:ctm3}. The distribution of estimated exposures shown in the map of median values (of the posterior distributions) shown in Figure \ref{map:china1} can also be seen in Figure \ref{map:milliegram1}  which the profile of air pollution (PM$_{2.5}$) in this country contains three distinct components: (i) a land mass with low levels of air pollution; (ii) a much larger proportion of the total land mass with (comparatively) high levels; and (iii) a substantial area with very high levels. In terms of potential risks to health, it is high levels in areas of high population that will drive the disease burden. Figure \ref{map:milliegram2} shows the distribution of estimated population level exposures, calculated by multiplying the estimate in each grid cell by its population. It can be seen that only a small proportion of the population reside in areas with the lowest concentrations  with the vast  majority of the population experiencing much higher levels of PM$_{2.5}$.

\section{Discussion}

In this paper we have developed a model to produce  a comprehensive set of high-resolution estimates of exposures to fine particulate matter. The approach builds on that used for the GBD2013 project that calibrated ground measurements against estimates obtained from satellites and a chemical transport model using linear regression.  This allowed data from the three sources to be utilised, but only provided an informal analysis of the uncertainty associated with the resulting estimates of exposure. There was also limited scope for considering changes in the calibration functions  between  geographical regions. As discussed in \citet{brauer2015exposure}, the increase in the availability of  ground measurements  has increased the feasibility of allowing spatially varying calibration functions. This was performed using geographically weighted regression in \citet{van2016global} but here  a hierarchical modelling approach is used in which country-specific calibration functions are used and information `borrowed' from the surrounding region and super--region where local  monitoring data is inadequate for stable estimation of the coefficients in the calibration models.  This is achieved using sets of random effects, for countries within regions within super--regions, reflecting a nested geographical hierarchy. The models are fit within a Bayesian hierarchical framework which produces (approximations to) full posterior distributions for estimated levels of PM$_{2.5}$ for each grid-cell rather than just point estimates. Summaries of these posterior distributions can be used  to give point estimates, e.g. mean, median,  measures of uncertainty, e.g. 95\% credible intervals, and exceedance probabilities, e.g. the probability of exceeding air quality guidelines.  Based on  posterior estimates (medians for each grid cell),  it is estimated that 92\% of the world's population reside in grid cells for which the annual average is  greater than the WHO guideline of 10 $\mu$gm$^{-3}$, which is greater than the 87\% reported in \citet{brauer2015exposure} for 2013. \\

In addition to the hierarchical approach to modelling, there has been increased availability of   ground monitoring data with measurements from 6003 locations (compared with 4073 for GBD2013) and estimates of specific components of air pollution, including mineral dust and the sum of sulphate, nitrate, ammonium and organic carbon, were available from atmospheric models. A series of candidate models, containing different sets of variables and structures for the random effects, were considered with the final choice of model being made on  predictive ability. This was assessed by cross-validation in which   models were fit to 25 training datasets (each containing  80\%  the overall data with stratified sampling to ensure samples were representative in terms of the distribution of concentrations within each super--region) and predictions compared to measurements within the corresponding validation set. The final model contained information on local network characteristics, including whether PM$_{2.5}$ was measured or values converted from PM$_{10}$, and whether the exact site type and location were known, together with satellite based estimates,  estimates of specific components from the GEOS-Chem chemical transport model,  land use and elevation, and population.  The final model includes country (within region, within super--region) random effects for satellite based estimates  and neighbouring country level random effects for population, with interactions between the fixed effects for  variables and those reflecting local network characteristics. Notably, the estimates of PM$_{2.5}$ from the TM5-FASST model used in GBD2013 were not found to improve the predictive ability and they were not included in the final model. In preference, estimates of specific components of pollution and the interaction between altitude and land--use from \citet{van2016global} were found to provide marked improvements in  predictively ability and are included in the model.  \\

The model presented here has been shown to offer improved estimates of PM$_{2.5}$ but there is certainly room for improvement, especially in areas  such as  Sub-Saharan Africa and South Asia. One of the potential uses of the outputs from the model, i.e. the information on areas with high predicted exposures and high uncertainty shown  in  Figures \ref{map:world} and \ref{map:china}, would be to guide where  future monitoring efforts might be focused. It may also be possible to utilise other sources of information related to air quality in addition to those considered here, such as road networks and other land--use variables. However, integrating data from many different sources, each of which will have their own error structures and  spatially varying biases, may require a different approach than the calibration one used here.  Bayesian melding provides a coherent framework in which   data from different sources at different levels of aggregation can be integrated, and allows for prediction at any required level of aggregation with associated estimates of uncertainty. However, in its current incarnation using MCMC, it is computationally infeasible  for large-scale problems of this type. Future research will involve developing  computationally efficient methods for performing Bayesian melding,  using approximate  Bayesian inference. \\
 
 \section{Concluding remarks}
 
In September 2015, world leaders set a target within the Sustainable Development Goals of substantially reducing the number of deaths and illnesses from air pollution by 2030. Following this, in May 2016, the World Health Organization approved a new ‘road map’ for accelerated action on air pollution and its causes.\\

The model developed here presents an important step forward in the understanding of global air pollution. It successfully integrates data from many sources to provide information on population-weighted exposures to fine particulate matter, one of the pollutants most strongly associated with adverse effects on human health. The result is comprehensive coverage of estimates of exposures for all countries, including areas where ground monitoring may be sparse. In addition to providing the most accurate estimates of global exposures to air pollution to date, the model provides valid estimates of uncertainty associated with those exposures. This results in a wealth of information on levels of air pollution around the world and highlights areas within countries that exceed WHO air quality limits. Such information is vital for health impact assessment, policy support and developing mitigation strategies.

\section{Acknowledgments}

The model was developed by the a multi-disciplinary group of experts established as part of the recommendations from the first meeting of the WHO Global Platform for Air Quality, Geneva, January 2014. The resulting \emph{Data Integration Task Force} consists of authors 1, 4-9 and 12-16 of this paper together with members of the WHO (authors 10,11 and 17).  The views expressed in this article are those of the authors and they do not necessarily represent the views, decisions or policies of the institutions with which they are affiliated. The model  presented and reviewed at the second meeting of the Global Platform for Air Quality, Geneva, August 2015. Matthew Lloyd Thomas is supported by a scholarship from the EPSRC Centre for Doctoral Training in Statistical Applied Mathematics at Bath (SAMBa), under the project EP/L015684/1. Amelia Jobling was supported for this work by WHO contracts APW 201255146 and 201255393. 

\bibliographystyle{mattnat}

\bibliography{references.bib}

\end{document}